\documentclass[prb,showpacs,twocolumn,superscriptaddress,aps,a4paper]{revtex4-1}
\usepackage{float}
\usepackage{dcolumn}
\usepackage{amsmath}
\usepackage{graphicx}
\usepackage{latexsym}
\usepackage{amsfonts}
\usepackage{amssymb}
\usepackage{bm}
\usepackage{color}

\DeclareGraphicsExtensions{.pdf,.gif,.jpg}

\newcommand{\be}{\begin{equation}}
\newcommand{\ee}{\end{equation}}
\newcommand{\bea}{\begin{eqnarray}}
\newcommand{\eea}{\end{eqnarray}}

\def\d{\delta}

\def\r{\rho}

\def\t{\tau}

\def\w{\omega}
\def\W{\Omega}

\def\blr{{\mathbf r}}

\def\bra{\langle}
\def\ket{\rangle}

\def\1op{\hat{\mathbbm{1}}}
\def\1{\mathbbm{1}}

\begin{document}

\title{Ultrafast Charge Migration in XUV Photoexcited 
Phenylalanine: a First-Principles Study Based on
Real-Time Nonequilibrium Green's Functions}

\author{E. Perfetto}

\affiliation{CNR-ISM, Division of Ultrafast Processes in Materials (FLASHit),
Area della Ricerca di Roma 1, Via Salaria Km 29.3, I-00016 Monterotondo Scalo, Italy}
\affiliation{Dipartimento di Fisica, Universit\`{a} di Roma Tor Vergata,
Via della Ricerca Scientifica 1, 00133 Rome, Italy}

\author{D. Sangalli}

\affiliation{CNR-ISM, Division of Ultrafast Processes in Materials (FLASHit),
Area della Ricerca di Roma 1, Via Salaria Km 29.3, I-00016 Monterotondo Scalo, Italy}

\author{A. Marini}

\affiliation{CNR-ISM, Division of Ultrafast Processes in Materials (FLASHit),
Area della Ricerca di Roma 1, Via Salaria Km 29.3, I-00016 Monterotondo Scalo, Italy}

\author{G. Stefanucci}

\affiliation{Dipartimento di Fisica, Universit\`{a} di Roma Tor Vergata,
Via della Ricerca Scientifica 1, 00133 Rome, Italy}
\affiliation{INFN, Sezione di Roma Tor Vergata, Via della Ricerca
Scientifica 1, 00133 Rome, Italy}
\email{gianluca.stefanucci@roma2.infn.it}





\begin{abstract}
The early stage density oscillations 
of the electronic charge in molecules irradiated 
by an attosecond XUV pulse takes place on femto- or subfemtosecond 
timescales. This ultrafast charge migration process is a central topic in 
attoscience as it dictates the relaxation pathways of the 
molecular structure. A predictive quantum theory of ultrafast 
charge migration should 
incorporate the atomistic details of the molecule,  electronic 
correlations and the 
multitude of ionization channels activated by the broad-bandwidth 
XUV pulse.
In this work we 
propose a first-principles Non Equilibrium Green's Function  method 
fulfilling all three requirements, and apply it to 
a recent experiment on the photoexcited 
phenylalanine aminoacid. 
Our results show that dynamical correlations are necessary
for a quantitative overall agreement with the experimental data.
In particular, we are able to capture the 
transient  oscillations at frequencies $0.15$~PHz 
and $0.30$~PHz in the hole density of the amine group,
as well as their suppression and the concomitant development   
of a new oscillation at frequency $0.25$~PHz after about 
14 femtoseconds.
\end{abstract}

\maketitle


Photoinduced charge transfer through molecules is the 
initiator of a large variety of chemical and 
biological 
processes~\cite{RevModPhys.81.163,GI.2014,SLMPS.2016,RTT.2017,Nisoli-review}.
Remarkable examples are the charge separation  
in photosynthetic centers, photovoltaic blends and
catalytic triads or the radiation-induced damage of biological
molecules.
These phenomena occur on timescales of several 
tens of femtoseconds to picoseconds and, in general, 
the coupling of electrons to nuclear motion cannot  
be discarded~\cite{may2008,Rozzi13,Falke14}. 
However, the density oscillations of the electronic charge 
following an attosecond 
XUV pulse precedes any structural
rearrangement and 
takes place on femto- or subfemtosecond timescales. This early stage
dynamics is mainly driven by electronic 
correlations~\cite{KuleffCederbaum} and it is 
usually referred to as ultrafast {\it  charge migration}.
Charge migration dictates the relaxation pathways of the molecule, 
e.g., the possible fragmentation channels of 
the cations left after ionization~\cite{Calegari336}.
Understanding and controlling this early stage 
dynamics has become a central topic in ultrafast 
science~\cite{Krausaab2160,Yuan2017} as it would, in principle, 
allow us to influence the ultimate fate of the 
molecular structure.

The sub-femtosecond electron dynamics in photoexcited or photoionized 
molecules can be probed in real-time
with a number of 
experimental techniques, e.g., high-harmonic spectroscopy~\cite{drescher2002time}, 
laser streaking photoemission~\cite{Schultze1658} or (fragment) 
cation chronoscopy~\cite{Calegari336,uiberacker2007attosecond,Belshaw2012}.
On the theoretical side, the description of ultrafast
charge migration in attosecond XUV ionized molecules is a complex 
problem since the parent cation is left in a coherent superposition of 
several many-electron states~\cite{Nisoli-review}. In fact, the XUV-pulse bandwidth is as large 
as tens of eV, thus covering a wide range of ionization thresholds.
The resulting oscillations of the  
charge density do therefore depend in a complicated manner 
on the electronic structure of the molecule
and on the profile parameters of the 
laser pulse (intensity, frequency, duration).

Understanding ultrafast charge migration at a fundamental 
level inevitably requires a time-dependent (TD) quantum framework 
able to incorporate the atomistic details of the molecular structure. The numerical 
solution of the TD Schr\"odinger equation (SE) in the subspace of 
carefully selected many-electron states is certainly feasible 
for atoms but it becomes prohibitive already for diatomic molecules. 
Fortunately, many physical observables require only 
knowledge of the TD  charge density $n(\blr,t)$ [or the single-particle density 
matrix $\r(\blr,\blr',t)$] rather than the full  many-electron wavefunction.
Density functional theory (DFT) and its TD 
extension~\cite{RungeGross:84,Ullrich:12} are 
first-principles methods having $n(\blr,t)$ as basic variable. TD-DFT 
calculations scale linearly with the number of electrons (against the 
exponential scaling of configuration interaction calculations) and it 
has been successfully applied to study ultrafast
charge migration during and after 
photoionization in a number of 
molecules~\cite{Wopperer2017,DeGiovannini,Kus2013,ayuso2017,lara2016,Bruner2017,Choasap}.
Although TD-DFT is an exact reformulation of the TD-SE, in 
practice all simulations are carried out within the {\em adiabatic 
approximation}, i.e., by 
using the equilibrium DFT exchange-correlation (xc) potential 
evaluated at the instantaneous density $n(\blr,t)$. The adiabatic 
approximation lacks of {\em dynamical exchange-correlation} 
effects which often play a major role in 
charge migration processes. For instance, 
ultrafast charge migration  in different polypeptide molecules
would not be possible 
without dynamical correlations.\cite{CEDERBAUM1999205,Remacle02052006,Hennig2005} 
Double (or multiple) 
excitations~\cite{Neepa-double-ex,RevModPhys.80.3} and
ionizations~\cite{PUvLS.2015},
long-range charge transfer 
excitations~\cite{GritsenkoBaerends,MaitraTempel,FuksMaitra:14},
image-charge quasi-particle 
renormalizations~\cite{PhysRevLett.97.216405,PhysRevLett.102.046802,PhysRevB.85.075105} and  
Auger decays~\cite{PhysRevB.86.045114} are  
other processes 
missed by the adiabatic approximation. 

In this paper we take a step forward and propose a first-principles
approach that shares with TD-DFT the favourable (power-law) scaling 
of the computational cost with the system size but, at the same time, 
allows for the inclusion of dynamical 
correlations 
in a {\em systematic} and {\em self-consistent} manner. 
The approach is based on Non Equilibrium Green's 
Functions (NEGF)
theory~\cite{kadanoff1962quantum,danielewicz1984quantum,svl-book,Stan2009,balzer2012nonequilibrium,PhysRevB.34.6933,PUvLS.2015,LPUvLS.2014} 
 and at its core is the (nonlinear) equation of motion 
for the single-particle density matrix $\r$ in the Kohn-Sham (KS) basis. 

\begin{figure}[tbp]  
\includegraphics*[width=.4\textwidth]{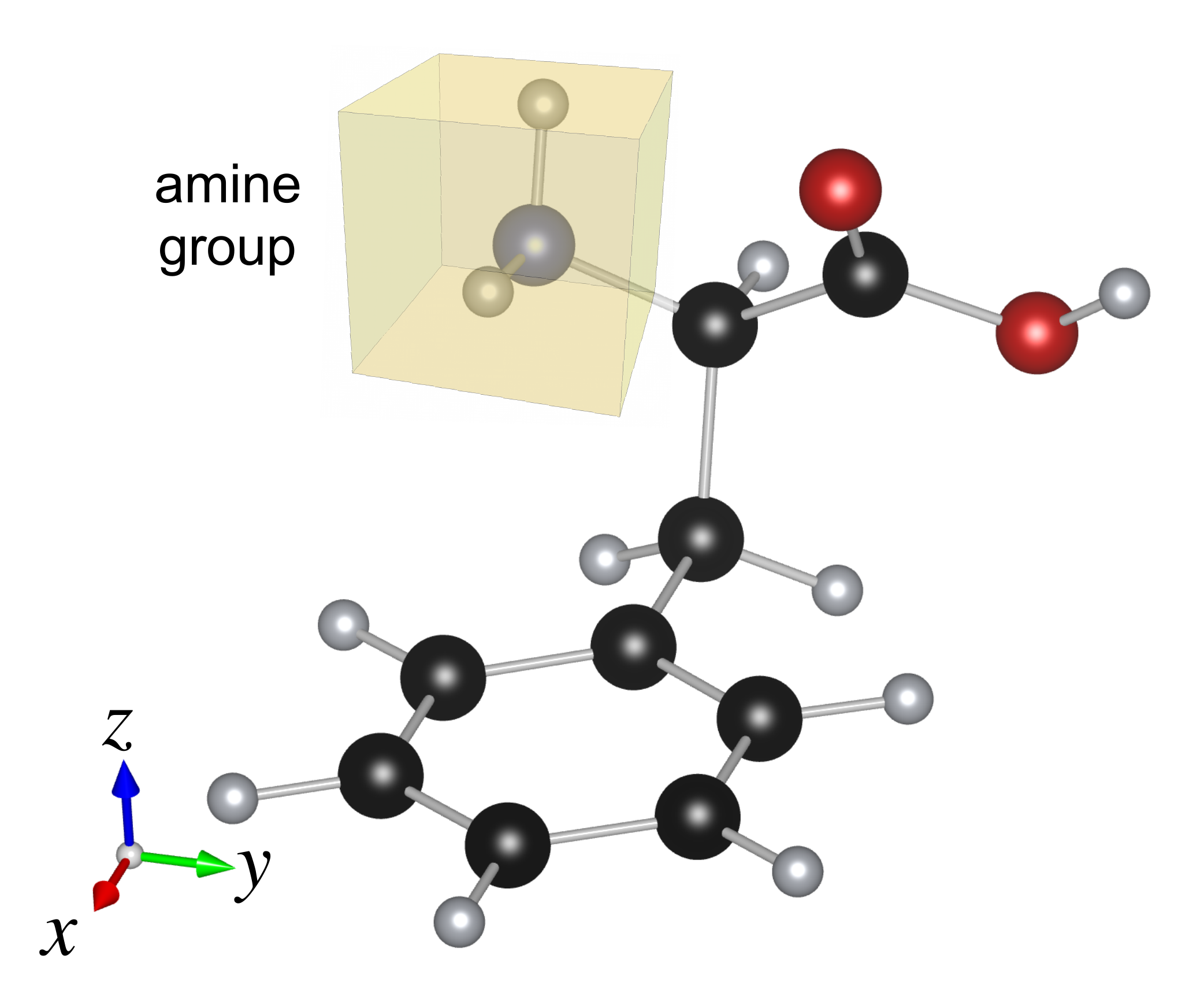}  
\caption{Molecular structure of the most abundant conformer of 
the phenylalanine molecule, see Ref.~\citenum{Calegari336}. 
Black spheres represent carbon atoms; 
gray spheres, hydrogen atoms; 
blue sphere, nitrogen; 
and red spheres, oxygen.
The amine group NH$_{2}$ is contained in a light-yellow cubic box.}  
\label{Fig_mol}  
\end{figure} 


The NEGF method is applied to revisit and complement the TD-DFT 
analysis of ultrafast charge migration in the 
phenylalanine aminoacid reported in Ref.~\citenum{Calegari336}, see 
Fig.~\ref{Fig_mol} for an illustration of the molecular structure.
In the experiment the ultrafast electron motion is activated by an 
ionizing XUV $300$-as pulse and it is subsequently probed 
by a VIS/NIR pulse. The probe causes a second ionization of the 
phenylalanine that eventually undergoes a fragmentation reaction. 
The yield $Y(\t)$ of immonium {\em dications} is recorded for different 
pump-probe delays $\t$.
The data show that for $\t\lesssim 14$~fs  the yield oscillates with a dominant
frequency $\W^{\rm exp}_{0}\approx 0.14$~PHz ($1$~PHz $=10^{15}$~Hz)  and 
a sub-dominant one $\W^{\rm exp}_{2}\approx 0.3$~PHz. For 
$\t\gtrsim 14$~fs, instead,  $Y(\t)$ 
oscillates almost monochromatically
at the frequency $\W^{\rm exp}_{1}\approx 0.24$~PHz. Ref.~\citenum{Calegari336} and 
other works~\cite{lehr2005} suggest that the yield of immonium dications 
shares common features with the TD hole density on the amine group 
NH$_{2}$, see Fig.~\ref{Fig_mol}, driven by the action of the XUV only.
This relation is also suggested  
by other experimental studies on $2-$phenylethyl$-N,N-$dimethylamine
and it is based on the hypothesis that to form an immonium dication 
the probe pulse is
absorbed by electrons on the NH$_{2}$.
The TD-DFT calculation 
of Ref.~\citenum{Calegari336} partially confirms this hypothesis, finding 
that the NH$_{2}$ hole density oscillates mainly at frequency $\W_{2}^{\rm DFT}\approx 
0.36$~PHz for $\t\lesssim 14$~fs and  $\W_{1}^{\rm DFT}\approx 
0.25$~PHz for $\t\gtrsim 14$~fs. 
However, in addition to the mismatch between $\W_{2}^{\rm exp}$ and $\W_{2}^{\rm DFT}$,
no clear evidence of the slow dominant oscillation at frequency 
$\W_{0}^{\rm exp}$ was found.

The main finding of this work is that
the inclusion of dynamical correlations 
through the proposed NEGF approach
is crucial to achieve a quantitative overall agreement with the experimental 
data. In particular, dynamical correlations are responsible for 
the appearance of a dominant oscillation at frequency 
$\W_{0}\approx 0.15$~PHz, for the renormalization of the high frequency 
$\W_{2}$ ($\approx 0.34$~PHz in mean-field and $\approx 0.30$~PHz in 
NEGF), and for the transition at delay $\t\approx 14$~fs from  
bichromatic to monochromatic behavior with frequency $\W_{1}\approx 
0.25$~PHz.

We examine  the most abundant conformer of the 
aminoacid phenylalanine~\cite{Calegari336} which
consists of a central CH unit linked to an amine
group (--NH$_{2}$), a carboxylic group (--COOH) and a benzyl group 
(--CH$_{2}$C(CH)$_{5}$).
We consider a linearly polarized 
XUV pulse with a weak peak intensity $I_{\rm pump}=5\times 
10^{11}$~$\mathrm{W/cm^{2}}$ (ensuring a linear response behavior), 
central photon energy $\w_{\mathrm{pump}}=30$~eV 
and duration $\t_{\mathrm{pump}} = 300$~as (with a $\sin^{2}$ envelope) 
yielding photon energies in the range $\sim (15,45)$~eV. 
Since the phenylalanine molecules of the
experimentally generated plume
are randomly oriented we perform calculations
for light polarization along the $x,y,z$ directions and average the 
results.
The XUV-induced ionization and the subsequent ultrafast charge migration are 
numerically simulated using the CHEERS$@$Yambo 
code~\cite{cheersyambonote,PS-cheers} that
solves the NEGF equation for the single-particle density matrix 
$\rho(t)$ in KS basis at fixed nuclei.  
Multiple ionization channels are taken into account by an exact 
{\em embedding} procedure for the KS continuum states while
dynamical correlations enter through a collision integral, which is a 
functional of $\r$ at all previous times.
Details on the theoretical method and numerical implementation are provided in 
the Supporting Information. 

\begin{figure}[tbp]  
\includegraphics*[width=.4\textwidth]{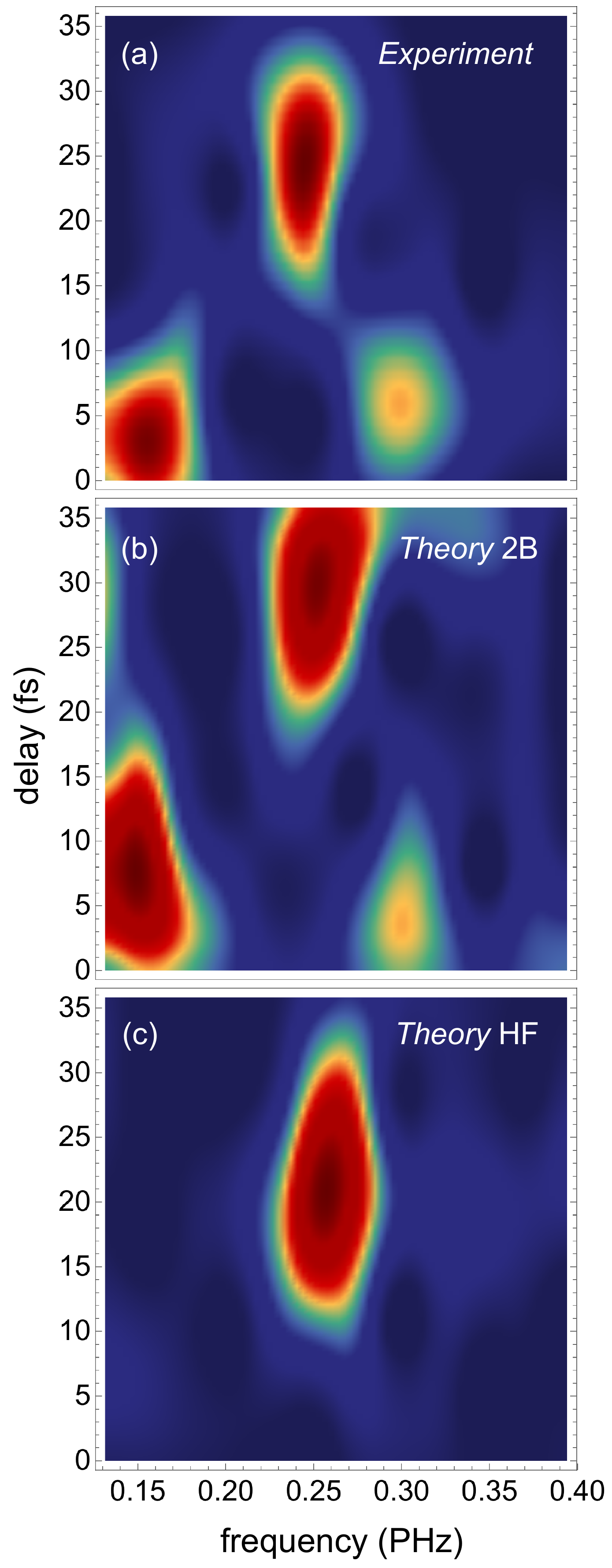}  
\caption{Spectrograms of 
(a) experimental yield $Y(\t)$ of immonium dications 
(row data from Fig.~S4 of Ref.~\citenum{Calegari336}) (b-c)
variation $\d N_{\rm amine}(t)$ of the charge on the amine group 
calculated within the 2B approximation (b) and HF approximation (c). 
All spectrograms are obtained by performing a Fourier 
transform with a sliding gaussian window-function of width $10$~fs 
centered at delay $\t_{d}$ (vertical axis), see Eq.~(\ref{spectrograms-eq}).
}  
\label{sliding}  
\end{figure}

To highlight the role of dynamical correlations we solve 
the NEGF equation  in the (mean-field) Hartree-Fock 
(HF) approximation and in the (beyond mean-field) second Born (2B) 
approximation. The latter has been shown to be accurate
for equilibrium spectral properties~\cite{schuler2017} 
and total energies~\cite{DahlenLeeuwen2005} of several molecules.
More importantly for the present work, the 2B approximation 
faithfully reproduces the nonequilibrium 
behavior of finite and not too strongly correlated systems (like the 
phenylalanine molecule considered here).
This evidence emerges from benchmarks against numerically 
exact simulations in 1D atoms and molecules,~\cite{PhysRevA.82.033427} 
quantum wells,\cite{BalzerHermanns2012} weakly correlated 
Hubbard and extended Hubbard
nanoclusters,~\cite{Sakkinen-2012,HermannsPRB2014,CTPPBonitz2016,HopjanPRL2016,ReichmanEPL2016,Joost2017}
the Anderson model at finite bias~\cite{UKSSKvLG.2011} and 
photo-excited donor-acceptor tight-binding Hamiltonians.\cite{C60paper2018}
The fixed nuclei approximation is not expected to be   
too severe either (this is confirmed {\em a 
posteriori} by Fig.~\ref{sliding}). Considering the molecular 
structure in Fig.~\ref{Fig_mol}, the time-dependent variation of 
the electronic charge on the amine group is due to electron flow 
through the N--C bond. The N--C stretching mode is 
medium-weak and it has a period of about 25$\div$30~fs;
hence the electron dynamics up to $30$~fs is not too disturbed by 
this mode. Furthermore, nonadiabatic couplings become less important 
when the cationic wavepacket is a linear combination of several 
many-electron states 
spread over a broad energy range. This is precisely the situation of 
the experiment in Ref.~\citenum{Calegari336}
since the bandwidth of the ionizing XUV is as large as 30~eV.

%



After ionization, the coherent superposition of cationic 
states is characterized by several fs and sub-fs oscillations. 
Following the suggestion of Refs.~\citenum{Calegari336,lehr2005}
we have calculated the TD charge $N_{\mathrm{amine}}(t)$ 
on the amine group by integrating the electron density 
in the light-yellow box shown in Fig.~\ref{Fig_mol}. 
The box was carefully chosen to give the correct number of valence 
electrons in equilibrium, 
i.e., $N_{\mathrm{amine}}(0)=7$.
In analogy with the analysis of Ref.~\citenum{Calegari336} we perform a
sliding-window Fourier transform of the relative variation of 
$N_{\rm amine}(t)$ with respect to its time-averaged value 
$\bra N_{\rm amine}\ket$
\be
\tilde{N}_{\rm amine}(\t_{d},\w)=
\int \!dt\, e^{-i\w t}\,e^{-(t-\t_{d})^{2}/t_{0}^{2}}\,
\d N_{\rm amine}(t),
\label{spectrograms-eq}
\ee
where $\d N_{\rm amine}(t)\equiv N_{\rm amine}(t)-\bra N_{\rm 
amine}\ket$ and $t_{0}=10$~fs.
The resulting spectrograms are shown in Fig.~\ref{sliding} panel (b) 
(2B) and panel (c) (HF). The theoretical spectra 
are compared with the spectrogram
of the experimental yield in panel (a).  The latter has been 
obtained as in Eq.~(\ref{spectrograms-eq}) after  replacing  $\d 
N_{\rm amine}(t)$ with the yield $Y(t)$ taken from Ref.~\citenum{Calegari336}.

The agreement between the 2B spectrogram and the experimental one is 
astounding.
For $\t_{d}\lesssim 15$~fs they both exhibit two main 
structures almost at the same frequencies:  $\W_{0}\approx 0.15$~PHz and 
$\W_{2}\approx 0.30$~PHz (theory), 
$\W_{0}^{\rm exp}\approx 0.14$~PHz and $\W_{2}^{\rm exp}\approx 
0.30$~PHz (experiment). 
Remarkably, the 2B calculation reproduces 
the relative weight as well, the peak at lower 
frequency being much more pronounced.
At $\t_{d} \approx 15$~fs a bichromatic-monochromatic 
transition occurs, and again the 2B frequency $\W_{1}\approx 0.25$~PHz is 
very close to the experimental one  $\W^{\rm exp}_{1}\approx 0.24$~PHz.
Of course, as we are comparing the TD amine density with the TD yield 
of  immonium dications,  a 
quantitative agreement in terms of peak intensities, delays, etc. 
cannot be, in principle, expected.
Nonetheless, our results strongly  
corroborates the hypothesis of Ref.~\citenum{Calegari336} according to 
which 
the two quantities are tightly related.
\begin{figure}[tbp]  
\includegraphics*[width=.4\textwidth]{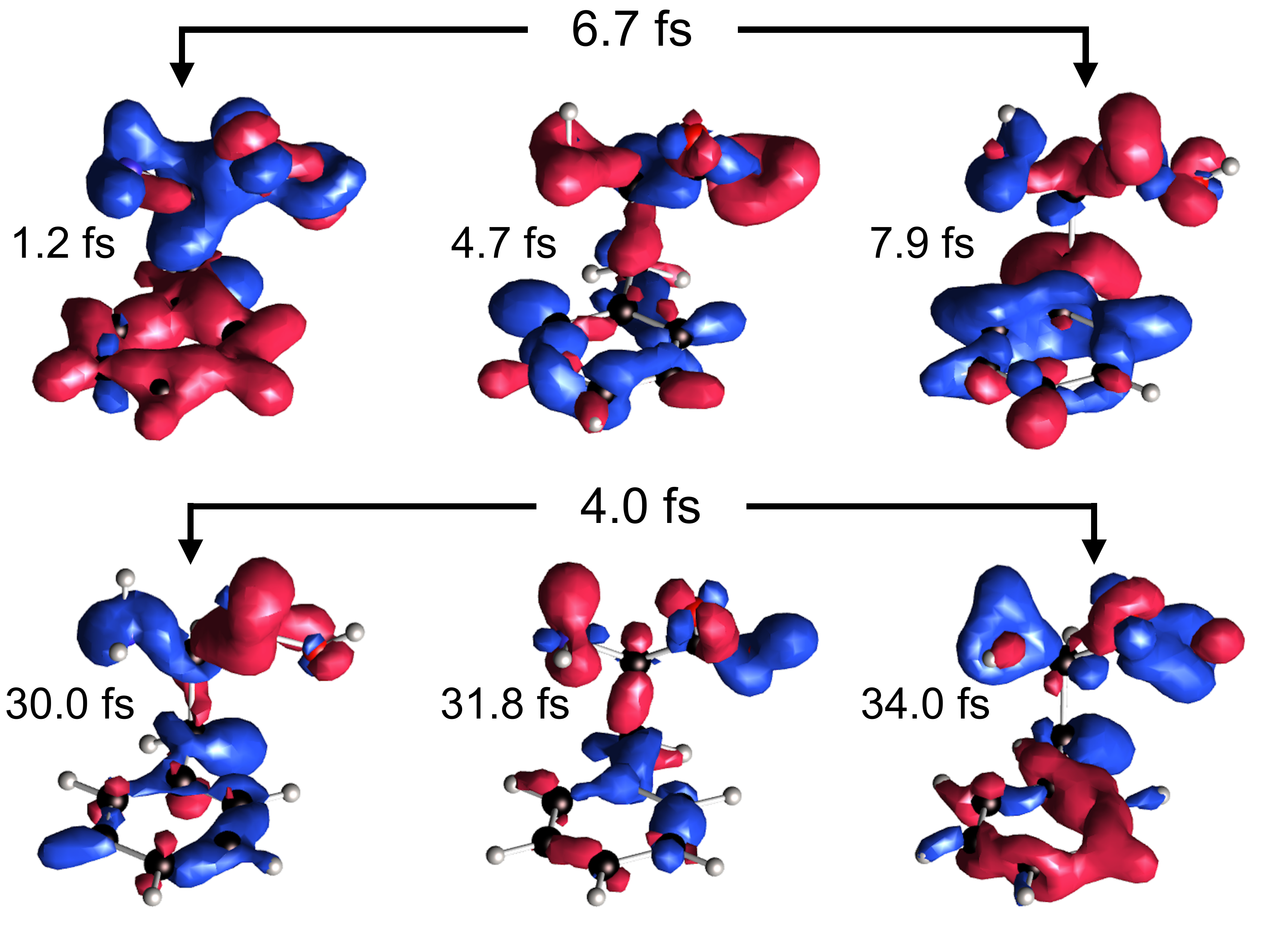}
\caption{Snapshots of the real-space distribution
of the molecular charge at three times corresponding to a maximum, consecutive minimum and then maximum
of $\d N_{\rm amine}(t)$ before (top) and after (bottom) the transition at 
$\t_{d} \approx 15$~fs. The snapshots highlight the two most prominent oscillations
at $\W_{0}\approx 0.15$~PHz (period $6.7$~fs) and 
$\W_{2}\approx 0.25$~PHz (period $4.0$~fs) observed in the correlated
spectrogram of Fig.~\ref{sliding} (b).
Hole excess (blue) and electron excess (red) are with respect 
to the reference density obtained by averaging $\rho(t)$ over the full 
real-time simulation. 
}  
\label{snaphots}  
\end{figure} 
In Fig.~\ref{snaphots} we display snapshots of the real-space distribution
of the molecular charge at 
three times corresponding to a maximum, consecutive minimum and then maximum 
of $N_{\rm amine}(t)$ before (top) and after (bottom) the transition at 
$\t_{d} \approx 15$~fs. The dominant oscillations at frequency 
$\W_{0}\approx 0.15$~PHz (period $6.7$ fs) and $\W_{1}\approx 
0.25$~PHz (period $4.0$ fs) are clearly visible. Interestingly, the 
periodic motion of the  charge on the amine group is not followed by other regions of the 
molecule.
This is a further indication of the role played by 
the quantity 
$N_{\rm amine}(t)$ in predicting 
the probe-induced molecular fragmentation.

The impact of dynamical correlations 
can be clearly appreciated in Fig.~\ref{sliding}~(c) where 
the spectrogram resulting from the mean-field HF approximation is 
shown. 
Overall, the agreement with the 
experimental spectrogram is rather poor.
We have a single dominant frequency $\W_{1}^{\rm HF} 
\approx 0.26$~PHz appearing at $\t_{d}\approx 12$ fs. 
Subdominant structures of frequencies $\W_{0}^{\rm HF} 
\approx 0.12$~PHz and $\W_{2}^{\rm HF} \approx 0.34$~PHz exist too but 
are not visible. 

\begin{figure}[tbp]  
\includegraphics*[width=.4\textwidth]{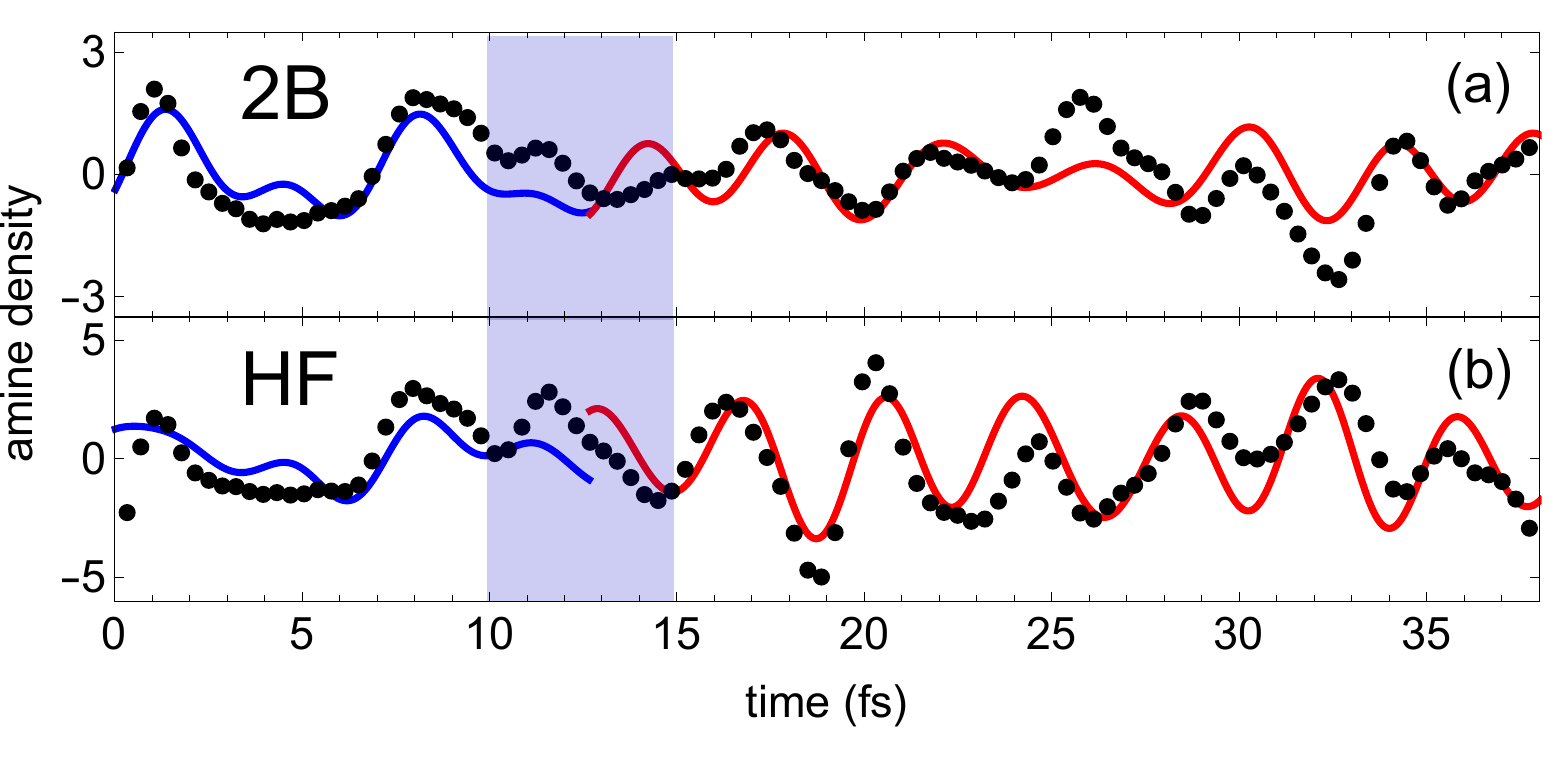}
\includegraphics*[width=.4\textwidth]{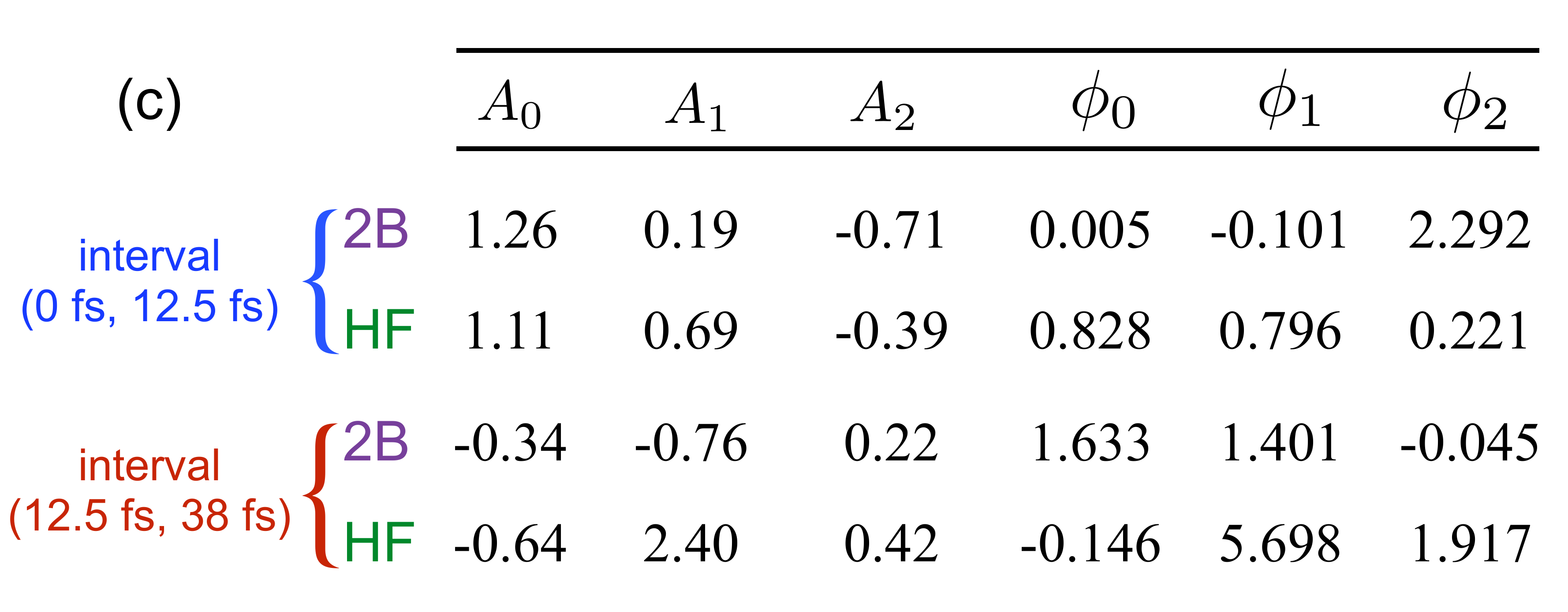}
\caption{
Panel (a) and (b): Relative variation $\d N_{\rm amine}(t)$ ($\times 
10^{6}$)
after filtering out all oscillations faster than $0.4$~PHz, obtained within the 2B and HF
approximations (dotted curves). Solid curves are trichromatic best fits 
obtained with the function $\sum_{i=0}^{2}A_{i}\sin(\W_{i}t+\phi_{i})$ in the 
time window  $(0\,\mathrm{fs},\,12.5\,\mathrm{fs})$ (blue) and
 $(12.5\,\mathrm{fs},\,38\,\mathrm{fs})$ (red).
Panel (c): Best fitted amplitudes $A_{i}$ (scaled by a factor $10^{6}$) 
and phases $\phi_{i}$}  
\label{tdamine}  
\end{figure}

In fact, the hight of the corresponding peaks is at least 2.5 times 
smaller  than that of the dominant one  
[see also 
Fig.~\ref{tdamine}~(c)]. In the Supporting Information 
we display the 3D plot of all three spectrograms 
to better appreciate the relative weight of the 
various structures. 


\begin{figure}[tbp]  
\includegraphics*[width=.4\textwidth]{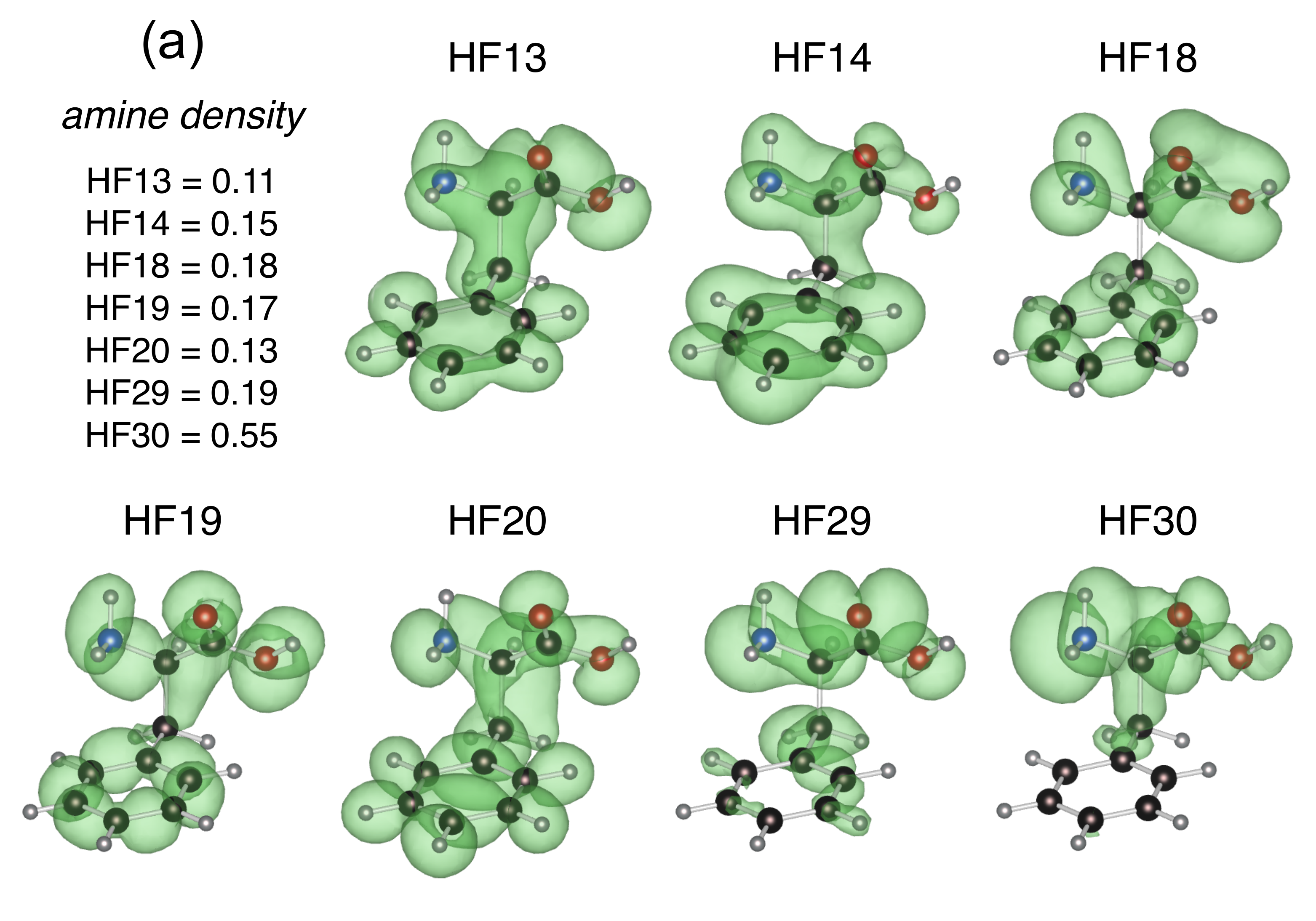}
\includegraphics*[width=.4\textwidth]{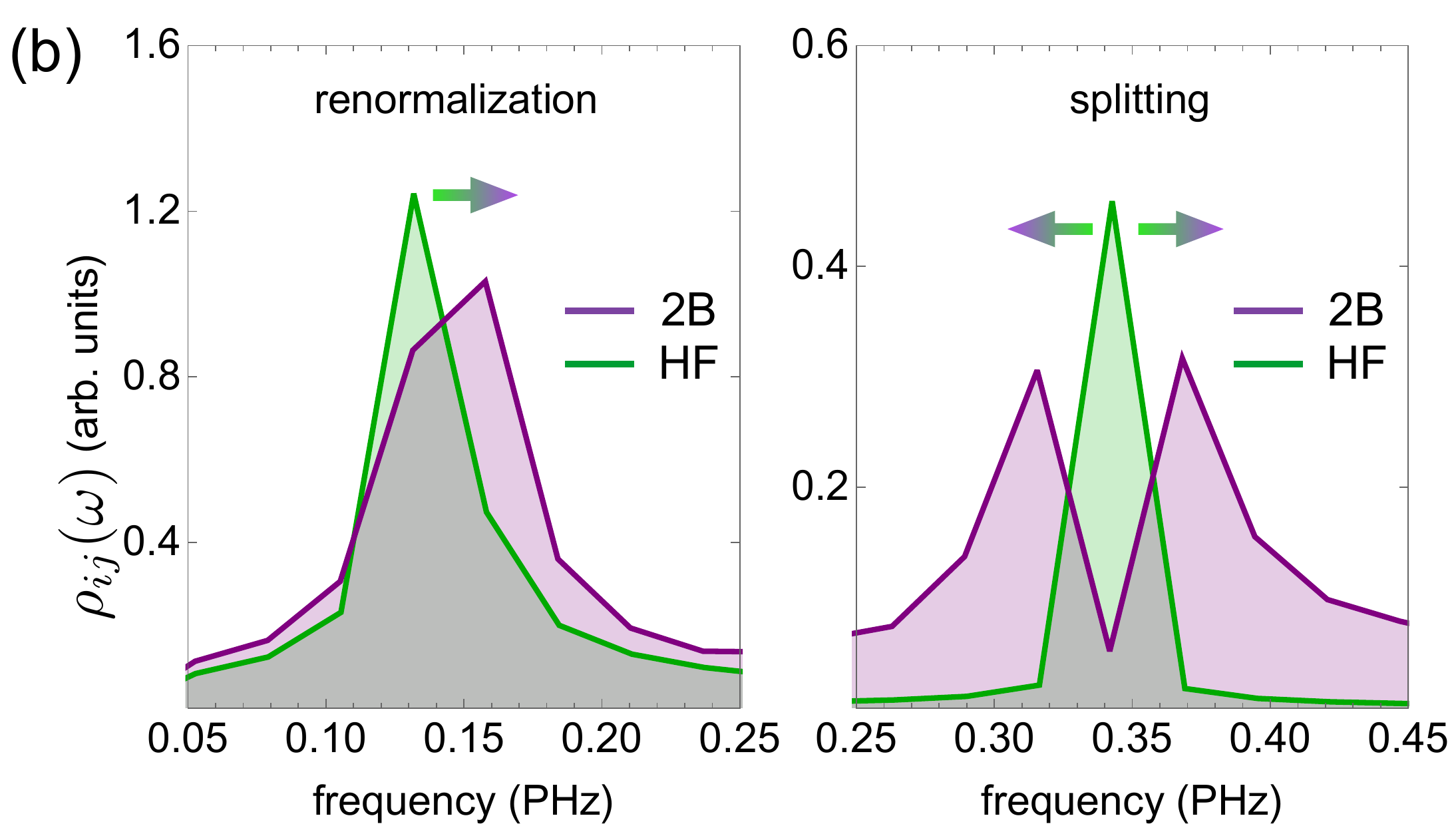}
\caption{Panel (a): Real-space plot of the HF orbitals with sizable 
charge on the amine group (see upper-left corner).
Panel (b): Fourier transform of selected elements of the 
single-particle density matrix $\rho_{ij}(t)$ in the HF basis for 
frequencies close to $\W_{0}$ (left) and $\W_{2}$ (right).
The low frequency $\W_{0}$ is due to three (almost) degenerate 
excitations, namely  HF13$\leftrightarrow$HF14, HF18$\leftrightarrow$HF19
and HF19$\leftrightarrow$HF20 and the left plot shows 
$\sum_{(i,j)}|\r_{ij}(\w)|$ with $(i,j)=(13,14),(18,19),(19,20)$.
The high frequency $\W_{2}$ is due to the excitation
HF29$\leftrightarrow$HF30 and the right plot shows $|\r_{29,30}(\w)|$.
To obtain $\r(t)$ we have performed calculations in the HF and 2B 
approximations for the three orthogonal 
polarizations of the XUV pulse 
and then we have averaged the results.
}  
\label{polamine}  
\end{figure} 

To get further 
insight into the dynamics of ultrafast charge migration, let us  
denote by HF$m$, $m=1,\ldots,32$, the HF orbitals of the 
64 valence electrons of the phenylalanine
 (ordered according to increasing values of the HF energy, 
hence HF32 is the HOMO). 
In NEGF calculations the HF basis is special since the 
HF orbitals are fully occupied or empty in the HF approximation 
and hence these are the reference orbitals to identify 
correlation effects like double or multiple excitations~\cite{svl-book}.
In Fig.~\ref{polamine}~(a) we have 
singled out those HF$m$  with a sizable 
amplitude on the amine group. 
We show the square modulus of their
wavefunctions along with the value (upper-left corner) 
of the respective spatial integrals over 
the light-yellow box of Fig.~\ref{Fig_mol}.
Figure~\ref{polamine}~(b)  
contains the Fourier transform of the single-particle density matrix 
$\rho_{ij}(t)$ for 
the most relevant excitations HF$i\leftrightarrow$HF$j$.
We display separately two spectral regions, one close to $\W_{0}= 
0.15$~PHz (left) and the other close to $\W_{2} \approx 0.30$~PHz 
(right). The results have been obtained using the 2B and HF 
approximations and by averaging over the three orthogonal 
polarizations of the XUV pulse (more details are provided in 
Supporting Information). 
The low frequency $\W_{0}$ is due to three (almost)
degenerate excitations, namely  HF13$\leftrightarrow$HF14, HF18$\leftrightarrow$HF19
and HF19$\leftrightarrow$HF20, and the left panel of 
Fig.~\ref{polamine}~(b)  shows the sum of them. In HF these 
excitations are slightly red-shifted, $\W_{0}^{\rm HF} = 0.12$~PHz, 
and the corresponding peak is hardly visible in the spectrogram of
Fig.~\ref{sliding}~(c). 
Dynamical (2B) correlations  redistributes substantially the  
spectral weight and give rise to a renormalization of about $0.03$~PHz 
($0.12$~eV), moving the low frequency much closer to the experimental value.
The high frequency $\W_{2}$ is due to the excitation
HF29$\leftrightarrow$HF30. The involved HF orbitals  are those with the largest amplitude on the 
amine group, in agreement with Ref.~\citenum{Calegari336}.
In the HF calculation the HF29$\leftrightarrow$HF30 excitation
occurs at $\W^{\rm HF}_{2}\approx 0.34$~PHz (TD-DFT predicts a 
slightly larger value 
$\approx 0.36$~PHz~\cite{Calegari336}) and, as the low frequency excitation, 
it is not detected by the spectrogram, see  
Fig.~\ref{sliding}~(c).
The effect of dynamical (2B) correlations is to split $\W^{\rm HF}_{2}$
into a doublet with 
$\W_{2}^{+} \approx 0.38$~PHz and $\W_{2} \approx 0.30$~PHz, 
suggesting that the underlying excitations are actually 
double excitations~\cite{Sakkinen-2012,svl-book}. 
Moreover, the redistribution of spectral weight 
makes visible only the  structure at $\W_{2}$, in excellent
agreement with the experimental spectrogram.
We mention that the central dominant frequency $\W_{1}=0.25$~PHz, 
see Fig.~\ref{sliding}, depends only weakly 
on electronic correlations since it occurs at almost the same energy 
and delay $\t_{d}$ in the 2B, HF and TD-DFT~\cite{Calegari336}.
In our simulations 
this frequency should be assigned to the HF18$\leftrightarrow$HF20 
excitation. See Supporting Information for more details.


We finally address the transition around $10 \div 15$~fs
leading to the suppression of the structures at $\W_{0}$ and 
$\W_{2}$ and the concomitant development of the central structure at $\W_{1}$.
We consider the 2B and HF curve $\d N_{\rm amine}(t)$
shown in Fig.~\ref{tdamine} (dotted) and 
perform two different trichromatic fits with 
the function $\sum_{i=0}^{2}A_{i}\sin (\W_{i}t+\phi_{i})$
in the time intervals $(0 \, \mathrm{fs},12.5 \, \mathrm{fs})$ 
and $(12.5 \, \mathrm{fs},38 \,\mathrm{fs})$, see blue and red curves respectively.
The fitting parameters are the amplitudes and phases whereas the 
frequencies are $\W_{i}$ in 2B, panel (a), and $\W_{i}^{\rm HF}$ in 
HF, panel (b).
The values of $A_{i}$ and $\phi_{i}$ 
are reported in panel (c).
We clearly see a dramatic change of the fitted amplitudes across the 
transition around $10 \div 15$~fs, consistent with the spectrograms 
of Fig.~\ref{sliding}.



In conclusion, we proposed a first-principles NEGF approach to study 
ultrafast charge 
migration during and after the action of an ionizing XUV pulse on the 
phenylalanine molecule.
Multiple ionization channels 
of the initially correlated many-electron system were 
 taken into account through an exact
{\em embedding} procedure for the KS states in the continuum.
Dynamical correlation effects were included 
at the level of the self-consistent second Born approximation, thus 
incorporating double excitations and other scattering mechanisms in the 
electron dynamics. 
The obtained results indicate that dynamical correlations 
(and hence memory effects) are crucial
to achieve a quantitative agreement with the experimental 
data. In fact, although the charge density oscillation
of frequency $\W_{1}$ at 
delays larger than $\t_{d}\approx 15$~fs is captured even in HF, the 
mean-field results do not display any significant 
structure at smaller delays. 
On the contrary,  the correlated NEGF 
calculations show a substantial reshaping,
characterized by the monochromatic-bichromatic 
transition $\W_{1}\leftrightarrow (\W_{0},\W_{2})$.
All frequencies as well as the delay of the transition are
in excellent agreement with the experiment.
The overall similarity between the theoretical and experimental 
spectrograms 
corroborates the existence of a 
tight relation between the charge on the amine group  and 
the yield of immonium dications. 

We finally observe that the NEGF approach proposed here can be extended  in at 
least two different ways to include the effects of nuclear 
motion
 The first is 
through the Ehrenfest approximation and it requires to update the one-particle 
and two-particle integrals during the time-propagation. The second 
stems from many-body perturbation theory and it consists in adding the 
Fan self-energy\cite{Fan-PhysRev.82.900} with equilibrium vibronic 
propagators to the electronic correlation self-energy. 
These developments allow for incorporating either classical effects 
or quantum harmonic effects, thus opening the door to studies of 
a broader class of phenomena.

\section*{Acknowledgments}

We thank P. Decleva for providing us with the nuclear coordinates of the
most abundant conformer of the phenylalanine molecule.
G.S. and E.P. acknowledge EC funding through the RISE Co-ExAN (Grant No. GA644076).
A.M., D.S., and E.P. also acknowledge funding from the European Union project 
MaX Materials design at the eXascale H2020-EINFRA-2015-1, Grant Agreement No.
676598 and Nanoscience Foundries and
Fine Analysis-Europe H2020-INFRAIA-2014-2015, Grant Agreement No. 654360.
G.S. and E.P.  acknowledge the computing facilities provided by the
CINECA Consortium within IscrC\_AIRETID and the INFN17$_{-}$nemesys project under the
CINECA-INFN agreement.



\begin{thebibliography}{62}
\bibitem [{che()}]{cheersyambonote}%
  \BibitemOpen
  \href@noop {} {}\bibinfo {note} {CHEERS@Yambo is an ongoing project stored in
  a dedicated private repository under the Yambo organization on GitHub, See
  Ref.~\citenum{MARINI20091392}.}\BibitemShut {Stop}%
\end{thebibliography}

%

\end{document}